**Revisiting 154-day periodicity in the occurrence of hard flares. A planetary influence?**

Ian Edmonds
12 Lentara St, Kenmore, Brisbane, Australia 4069.
Ph/Fax 61 7 3378 6586, ian@solartran.com.au

**Abstract.**

Rieger et al (1984) reported observations of a 154 day periodicity in flares during solar cycle 21. This paper discusses the observations in the light of a simple empirical planetary model of sunspot emergence. The planetary model predicts sunspot emergence when Mercury and Earth approach conjunction and Mercury approaches the Sun. We show that the reported times of flare activity are coherent with the planetary model. While the base period of the model is 170 days, the average model period, over the interval of flare recordings, is 157 days due to a 180 degree phase change in the planetary forcing near the middle of the record interval. We conclude that the periodicity at 154 days arises when the phase change in planetary forcing and the resulting progressive phase change in total sunspot area emergence and flare occurrence shifts the major peak in the flare spectrum from the planetary forcing period, 170 days, to 154 days.

**1. Introduction.**

The observation of an ~ 154 day periodicity in the occurrence of hard flares, Rieger et al (1984), hereafter R84, initiated a now 32 year long investigation of mid-range, 40 – 2000 day, periodicities in solar activity. R84 analysed 139 flares recorded by the gamma ray spectrometer on the Solar Maximum Mission (solar cycle 21) from February 1980 through to the end of 1983 when recording ceased. It was observed that the flares occurred in eight bursts spaced by about 5 months. This was determined by folding the flare-event times with a period of 154 days and observing that 35% of the flares occurred within a bin of 15.4 day width between phase 0.4 and 0.5 on the phase histogram. The times of phase 0.4 were 25/5/80, 27/10/80, 30/3/81, 1/9/81, 2/2/82, 6/7/82, 8/12/82 and 12/5/83 and these times can be used to identify the time of occurrence of each of the flare bursts, (e.g. Plate 9, Massi 2006). Figure 4 of R84 also showed that the major bursts of flare activity occurred at the times of secondary peaks in sunspot number data. In the following years researchers reported periodicities in many solar related indices with, however, an extreme focus on ~155 day periodicity. For example, Lean (1990), in a comprehensive survey of periodicity in sunspot number and sunspot areas, found dominant periodicity at 249 days and 353 days and relatively minor periodicity at 155 days. Nevertheless, Leans subsequent analysis focused exclusively on 155 day periodicity and the dominance of 249 day and 353 day periodicity was not mentioned in the abstract. See also (Bogart and Bai 1985, Bai and Cliver 1990, Ballester et al 2002, Bai and Sturrock 1993, Oliver et al 1998, Ballester et al 1999, Ballester et al 2002, and Ballester et al 2004). It is now known that a wide range of different periodicities occur intermittently in solar activity data, e.g. Tan and Cheng (2013), Chowdhury et al (2009). Rieger periodicities, as they are now commonly known, have also been observed in other stars, Massi et al (2005). However, there is currently no accepted explanation for Rieger periodicity.

The possibility that Rieger periodicities may be associated with planetary motion has been discussed in terms of the effects planets have on the Sun, e.g. tidal effects, Scafetta (2012), or effects on the motion of the Sun about the solar system centre of mass, Juckett (2003), Charvatova (2007).  A planetary influence on solar activity is a controversial concept as the physical effects of planets on the Sun are extremely small, De Jager and Versteegh (2005), Charbonneau (2013). However, there is evidence of solar flares occurring at times when planets are in conjunction, Hung (2007), and planetary models have been developed to explain the decadal and longer periodicities in solar activity, e.g. (Scafetta 2012, Abreu et al 2012).  Recently, an empirical model based on planet conjunctions was developed and applied with some success to predicting intermediate range periodicity in sunspot area emergence, Edmonds (2016b). This paper applies a simplified version of the empirical model to the flare observations by R84.  Section 2 details data and methods. Section 3 of the paper outlines the planetary model. Section 4 compares the model with the data of R84. Section 5 is a discussion and Section 6 a conclusion.

**2. Data and methods.**

The observational data includes the flare observations by R84 and the daily sunspot area observations on the North and South hemispheres of the Sun (SSAN and SSAS) available at http://solarscience.msfc.nasa.gov/greenwch/daily_area.txt . Planetary data include the daily the orbital radius of Mercury, $R_M(t)$, and the solar longitudes of Mercury and Earth, respectively $\theta_M(t)$ and $\theta_E(t)$, available at http://omniweb.gsfc.nasa.gov/coho/helios/planet.html .

As the time variation of sunspot emergence is a complex of numerous components band pass filtering was used to isolate the component of interest. The ~ 176 day components of sunspot area emergence were obtained by performing a FFT on the sunspot area data and using Fourier pairs in the frequency range between 0.00483 days$^{-1}$ and 0.00653 days$^{-1}$ (period range 207 days to 153 days)  to synthesize the ~ 176 day period components of sunspot area. In the following, band pass filtered data is denoted by using the central period as a suffix e.g. the ~ 176 day period component of sunspot area North is denoted 176SSAN.  Where data has been smoothed by, for example, a 365 day running average, the smoothed data is denoted by the suffix Snnn e.g. a 365 day running average of total sunspot area data would be denoted SSAT S365.

**3. Planetary model of sunspot emergence.**

The planetary model is based on the idea that rapid changes or pulses in tides on the Sun trigger the emergence of sunspots, (Hung 2007, Scafetta 2012, Tan and Cheng 2013, Edmonds 2016b). The tidal effect due to a planet of mass M and orbital radius R is proportional to $M/R^3$, Hung (2007).  The tides due to two planets add when the planets are in conjunction, i.e. at times when the two planets and the Sun are aligned. As Mercury has the fastest angular velocity relative to the Sun the occurrence of fast tidal pulses coincides with the times of Mercury - planet conjunctions. The time variation of a pulse can be estimated, for example in the case of a Mercury-Earth conjunction, from the alignment index for Mercury and Earth, $I_{ME}(t) = <\cos(\theta_M(t) - \theta_E(t))>$ where $\theta_M(t)$ and $\theta_E(t)$ are the daily values of the angles of solar longitude of Mercury and Earth, Hung (2007). The alignment index, $I_{ME}(t)$, can be approximated by

$$y_{ME}(t) = 1 + \cos(2\pi t/T_{ME} + 0.7744) \quad \text{units} \tag{1}$$

where $T_{ME}$ is the period of Mercury-Earth conjunction, 57.9387 days, and 0.7744 is a phase angle in radians referenced to $t = 0$ on January 01 1876, Edmonds (2016b).

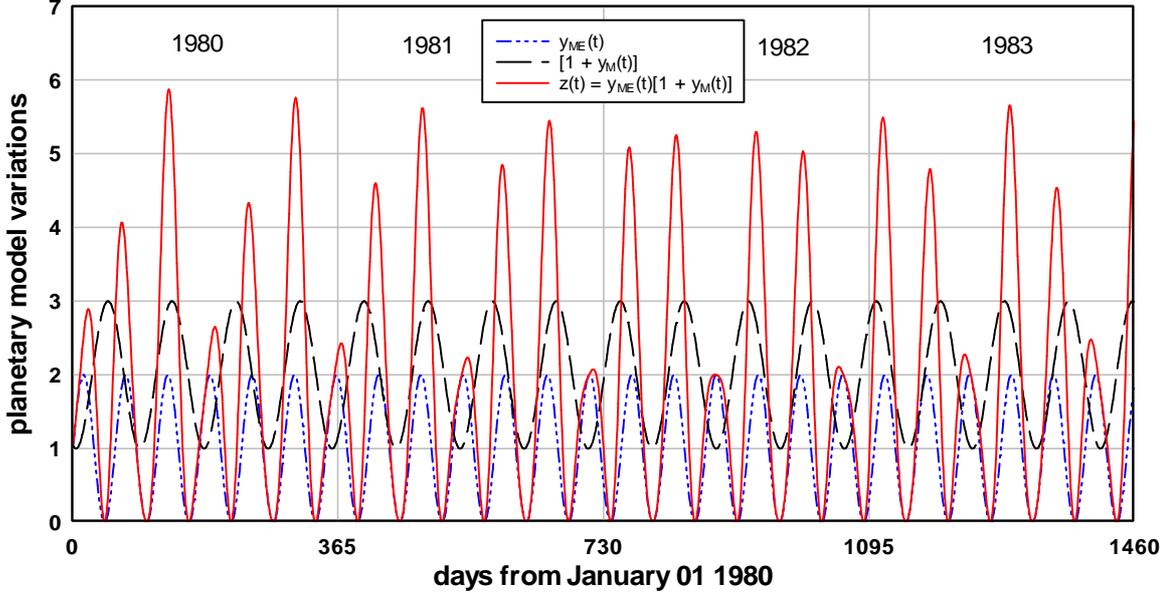

154 day periodicity model variations.grf
**Figure 1.** The alignment index for Mercury and Earth, $y_{ME}(t)$; the proportional variation of the Mercury tidal effect, $[1 + y_M(t)]$; and the modulation product, $z(t) = y_{ME}(t)[1+y_M(t)]$. The initial primary pulse of the model is overtaken in strength by the secondary pulse in 1982. As a consequence the model variation exhibits an $\sim \pi$ radian phase change in the primary pulse during 1982.

The daily variation of $y_{ME}(t)$, for 1980 to 1983, (day 37985 – 39445 relative to January 01, 1876) is shown in Figure 1. Notice that the alignment index varies from 0 when the angular difference between the planets is +/- 90°, and 2 when the angular difference is 0° or 180°. We expect the rate of change and the strength of the combined tidal pulse to be stronger if the conjunction occurs during closest approach of Mercury to the Sun. The proportional time variation of the Mercury tidal effect, $1/R_M(t)^3$, can be approximated by $1 + y_M(t)$ where

$$y_M(t) = 1 + \cos(2\pi t/T_M - 2.2553) \tag{2}$$

Here, $T_M = 87.969$ days is the orbital period of Mercury. The phase angle is relative to $t = 0$ on January 01 1876, the start of the continuous sunspot area record. The daily variation of $y_M(t)$ for 1980 to 1983 is also shown in Figure 1. We note that there is an approximately 1:3 variation in the proportional Mercury tidal effect between a minimum of 1 arbitrary unit and a maximum of 3 arbitrary units. The modulation of the planet conjunction effect with the Mercury orbital radius effect is given by

$$z(t) = [1 + y_M(t)]y_{ME}(t) \tag{3}$$

also shown in Figure 1. We notice that z(t) has a maximum of 6 arbitrary units when the conjunction term, $y_{ME}(t)$, is in phase with the Mercury proximity term, $[1 + y_M(t)]$. This gives rise to a primary pulse in the model. There are secondary and tertiary pulses between each primary pulse with the height of the tertiary pulse taking a minimum value of 2 units when the conjunction and proximity terms are in anti-phase. Note that the primary pulses are about 176 days apart. The primary pulse suffers an ~ π radian phase change when the secondary pulse overtakes the primary pulse in strength. This occurs, for example, during 1982, Figure 1. In this model an ~ π radian phase shift between primary pulses occurs at intervals of 2376 days, while between these intervals the primary pulse in the model variation is relatively stable. The model just described generates a spectral peak at period 169.7 days corresponding to the frequency difference $f_{ME} - f_M = 0.005892$ days$^{-1}$. Here we regard the model variation as representative of a planetary forcing of sunspot emergence. However, the physics of how this might occur is outside the scope of this paper. Although the relation $z(t) = [1 + y_M(t)]y_{ME}(t)$ is simple the resulting forcing is quite complex with abrupt phase changes occurring when the secondary pulse overtakes the primary pulse in strength. When the tertiary pulse overtakes the secondary pulse a similarly abrupt phase change occurs in the planetary forcing. Thus in this model abrupt phase changes occur at intervals of 2376/2 = 1188 days ~ 3.3 years. If other planet conjunctions were included in the model several other difference terms would be generated including $f_{MV} - f_M = 0.00247$ days$^{-1}$ (405 days) and $f_{MJ} - f_M = 0.00109$ days$^{-1}$, (91.7 days), $f_{MJ} - f_{MV} = 0.00845$ days-1 (118 days), Edmonds (2016b). However, here we focus on the Mercury – Earth conjunction as it generates intermediate range periodicity in the spectral region of the ~ 154 day periodicity described by R84.

**4. Comparison of solar activity with the planetary model during 1980 – 1983.**
**4.1 Hard flare occurrence.**
Figure 2 compares pulses in the planetary model, 30z(t), against the times of bursts of high energy flare activity. The times nominated by R84 correspond to the phase, 0.4, at the start of the phase bin between 0.4 and 0.5 that contained the highest fraction of flare bursts. As the bin was 15.4 days wide the reference lines marked in Figure 2 correspond to the R84 dates (listed above) plus eight days to bring the times to the centre of the phase bin. Also shown is the 365 day smoothed variation of total sunspot area, (SSAT/10 S365), during the same period. This variation indicates that the interval extends over the last half of solar cycle 21. We also include the model variation when the elliptical orbit of Mercury is taken into account. Evidently the approximations of equations 1 and 2 are minor. We note that the occurrence of flare bursts coincided closely in time with major pulses of the model variation (pulses > 100 units) during the maximum phase of the solar cycle. The correspondence was less evident for the last burst of flare activity as the solar cycle approached minimum.

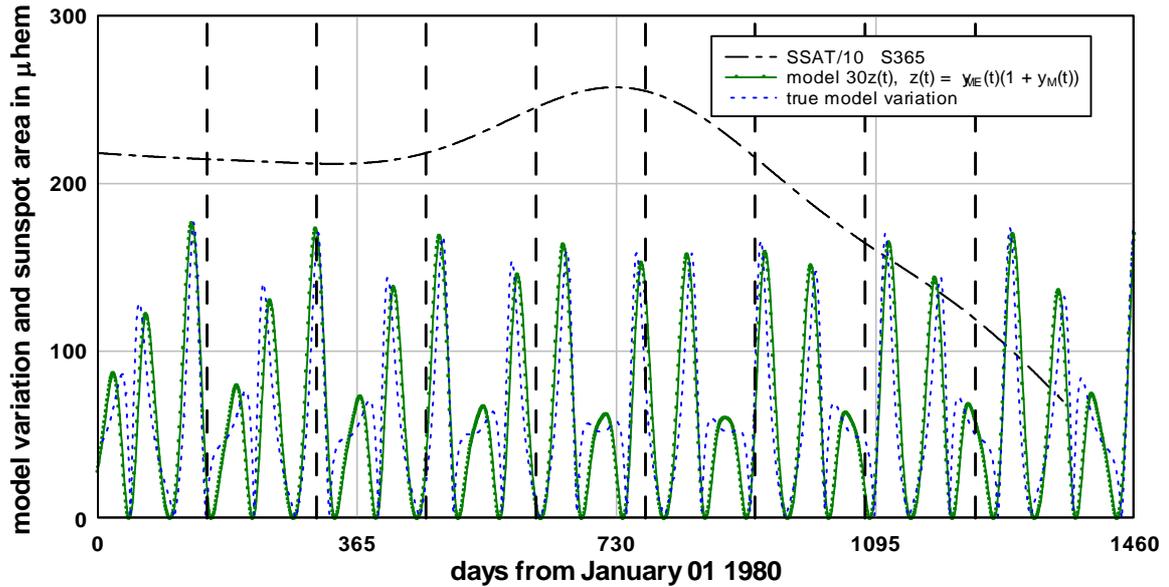

176SSAN 176SSAS 1980 81 82 83 and model.grf
**Figure 2.** The planetary model variation, $z(t) = 30y(t)$, with $y(t) = y_{ME}(t)[1 + y_M(t)]$ for 1980 to 1983. Also shown, the 365 day smoothed version of total sunspot area. Times of flare bursts, at the vertical reference lines, occur within 54 days of the primary pulses of the model. The true model variation, taking account of the elliptical orbit of Mercury is also shown.

We note that all of the marked times of flare bursts are within 54 days of the times of occurrence of the primary pulses of the model with the exception of the time of burst close to day 730 in Figure 2. In this case the flare burst time occurred almost on top of the time of the secondary pulse and 65 days from the following primary pulse. However, this is when a phase change in primary pulse occurs and the primary pulse is almost indistinguishable in strength from the secondary pulse so we ignore this exception and work on the basis that all eight flare bursts were within 54 days of a primary peak of the model. As the primary pulses are, except at phase change, 176 days apart the probability of a burst occurring by random chance within 54 days of a primary pulse is $p = 54/88 = 0.61$. The probability of this occurring by random chance in eight out of eight trials is $0.61^8 = 0.019$, about 2%. This result suggests that the primary pulses of the planetary model are influencing the occurrence of flares. This finding supports the observations of Hung (2007) that flares are more likely when one or more of the planets, (Mercury, Earth, Venus or Jupiter), are either overhead or underfoot a sunspot group.

**4.2 Sunspot area emergence.**
R84 also showed that the major bursts of high energy flares occurred near the times of peaks in sunspot number, (R84, Figure 4). In this section we compare the emergence of sunspot area on both the Northern hemisphere (SSAN) and on the Southern hemisphere, (SSAS) for the entire maximum of solar cycle 21, 1977 to 1984, (day 36890 – 39811 relative to t = 0 on January 01, 1876).

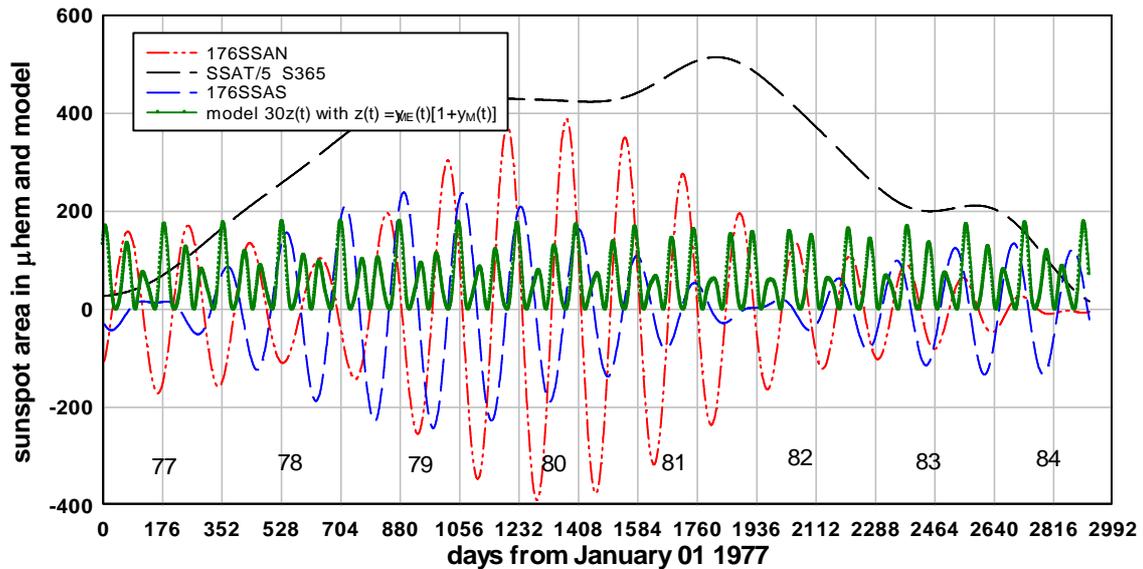

**Figure 3.** The ~176 day band pass filtered components of sunspot area North, (176SSAN), and South, (176SSAS), for solar cycle 21, 1977 to 1984, compared with the planetary model variation for the same interval. Also shown, the 365 day smoothed version of total sunspot area. Note that, initially, 176SSAS follows the primary pulse of the model whereas 176SSAN follows the secondary pulse of the model. Note that 176SSAS undergoes a π phase change at the beginning of 1982 when the primary pulse undergoes a π phase change, whereas, 176SSAN, which follows the secondary pulse as it becomes a primary pulse, does not experience a π phase change in 1982. The time axis is in intervals of 176 days to facilitate following the phase changes.

In Figure 3 the ~176 day components of SSAN and SSAS are compared with the model for sunspot emergence, $30z(t)$, with $z(t) = y_{ME}(t)[1 + y_M(t)]$. Consider first the 176SSAS component. After a phase change in 1977 the peak in 176SSAS follows the primary pulse of the model variation at a lag of about 15 days suggesting the primary pulse is influencing the emergence of Southern hemisphere sunspots. This situation continues through 1978, 1979, and 1980 when 176SSAS begins to respond to the increasingly strong secondary pulse, with which it is in anti-phase, by decreasing in amplitude to zero. In 1982, a phase change of ~ 180° occurs in the model variation with the secondary pulse of the model replacing the primary pulse in strength. The 176SSAS variation responds to the phase change by re-emerging at the end of 1982 to move increasingly into phase with the new primary peak of the model as the solar cycle approaches minimum in 1983 and 1984. That is, the 176SSAS component itself undergoes a phase change to bring it towards in-phase with new primary pulse of the model.

We consider now the variation in the 176SSAN component of sunspot emergence. From 1977 to 1978 176SSAN follows the secondary pulse of the model. In 1978 the tertiary pulse overtakes the secondary peak in strength and becomes the new secondary pulse. It appears that, from 1978, the 176SSAN component peak undergoes a phase change to follow the new secondary pulse of the model during 1978, 1979 and 1980. However, from 1981 the secondary model pulse begins to overtake the primary pulse in strength and in 1982 becomes the primary pulse. In response the 176SSAN variation continues to follow the secondary pulse as it replaces the primary pulse and, unlike the 176SSAS variation, the 176SSAN variation does not undergo a π phase change following the π phase change in the primary pulse in 1982.

The overall result of the 176SSAS variation following the model primary pulse and the 176SSAN variation following the model secondary pulse is that the 176SSAN and 176SSAS variations tend to be in quadrature during most of solar cycle 21 but come into phase in 1983 and 1984 towards the end of the solar cycle.

R84 compared total sunspot number with time of occurrence of flare bursts. The total sunspot area (SSAT) time variation is known to be closely similar to total sunspot number time variation. Here we compare the variation of 176SSAT with the times of flare bursts and the model variation during 1977 to 1984.

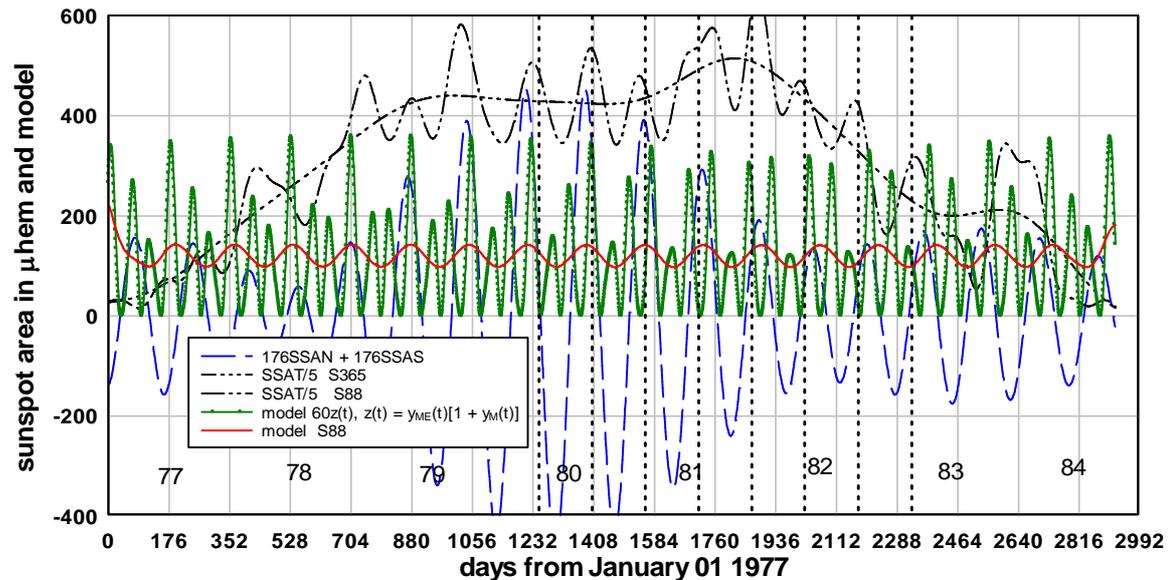

176SSAT and model 1980 to 1983.grf

**Figure 4.** The variation of the ~176 day band pass filtered component of total sunspot area (176SSAT = 176SSAN + 176SSAS) is compared with the planetary model variation and the 88 day smoothed version of the model. Also shown the 365 day and 88 day smoothed versions of total sunspot area (SSAT). The vertical reference lines mark the occurrence of flare bursts as observed by Rieger et al (1984). Note that all eight times of flare bursts occur within the positive excursions of 176SSAT indicating coherence of flare activity with this component of sunspot area and also coherence with the peaks in total sunspot area. The time axis is divided into 176 day intervals to facilitate following phase changes in the variations.

The times of bursts of flare activity during 1980 to 1983 as observed by R84 are indicated by the vertical reference lines in Figure 4. We note that all eight of these times fall within the intervals of maximum excursions of 176SSAT. As the peak to peak level of 176SSAT is ~ 800 μhem this component constitutes a significant fraction of the level of the 365 day smoothed (S365) variation of SSAT, ~ 2000 μhem at solar maximum. Thus it is likely that the peaks in sunspot number that R84 associated with flare bursts correspond to the peaks in total sunspot area associated with the 176 day component, Figure 4. We note that the observed correlation between sunspot area and flare activity during solar cycle 21 is consistent with the observation of the same correlation by Oliver et al (1998). We also note that the 176SSAT variation appears to be strongly correlated with the 88 day smoothed (S88) variation of the planetary model during solar cycle 21. For example, between January 01 1978 and December 31 1983, over thirteen 176 day cycles, the correlation coefficient between 176SSAT and the smoothed planetary model variation is +0.76. Over the entire record 1977 to 1984 the correlation coefficient is +0.55. This provides further support for the idea that sunspot emergence and flare activity is coherent with major pulses of the planetary model outlined above.

**5. Discussion.**

The result of comparing times of occurrence of flare bursts and times of occurrence of primary pulses of the planetary model, Figure 2, is straightforward. There is only 2% probability that the observed time coincidence could have occurred by chance.

Flares are a downstream result of sunspot emergence in the sense that when sunspots emerge periodically into the same active area on the Sun any subsequent interaction between the magnetic fields of new and old sunspots can lead, via reconnection, to the occurrence of flares, (Nishio et al 1997, Ballester et al 2002, Massi 2006). Thus periodicity in sunspot emergence should lead to periodicity in flares provided the sunspots emerge in the same region of the Sun. Edmonds (2016b) demonstrated that a planetary model of sunspot emergence with a base period of the 88 day Mercury orbital period should result in preferred longitudes of sunspot activity on the Sun. Thus the planetary model described above provides a link between two of the necessary conditions for periodic flare occurrence: periodic sunspot emergence and sunspot emergence at the same solar longitude. However, a third condition for flare occurrence is that the initial sunspot should persist long enough for interaction (reconnection) between its flux tube and the flux tube of the second sunspot to occur, Nishio (1997). The half life of sunspot groups is only a few days, Ringnes (1964), so very few sunspot groups would persist for 176 days with larger area sunspots being the most persistent, Henwood et al (2010). Although sunspots are much more numerous than flares the area occupied by sunspots is a very small fraction of the solar surface. Therefore, some mechanism of periodic emergence at the same longitude would increase the likelihood of flares occurring.

R84 obtained a power spectrum of flare occurrence with major peak at frequency 75 nHz (154 day period). So the question may be asked, why is the above analysis based on a planetary component at frequency $f_{ME} - f_M = 0.005892$ days$^{-1}$ = 68.1 nHz (170 day period) when a peak at this frequency does not appear in the power spectrum of flares reported by R84? The reason is that the planetary model generates a $\pi$ phase change in the primary pulse, at the beginning of 1982, ~ day 730 in Figure 2, which is near the middle of the record interval. This results in a $\pi$ phase change in 1982 to the 176SSAS variation which is following the primary pulse but results in no significant phase change in 176SSAN which is following the secondary pulse. As a result the component of total sunspot area emergence, 176SSAT, suffers a progressive phase change over the interval from the beginning of 1980 to the beginning of 1983. It is expected and is observed, see Figure 4, that flare occurrence is coherent with total sunspot area emergence and, equivalently, with sunspot number. Therefore we expect flare occurrence to also suffer a progressive phase change over the same interval. A progressive phase change either advances the periodic waveform, causing the peak of the waveform to occur progressively earlier, and leading to a shorter period over the record or retards the waveform, leading to a longer period over the record. It is evident from Figure 3 that 176SSAT, the sum of 176SSAS and 176SSAS, between 1980 and 1983 is progressively advanced in phase. That is, the peak of 176SSAT occurs progressively earlier in the 176 day cycle indicated on the time axis, and by 1983 has shifted by slightly more than π radians. As the base period of the planetary model is 170 days, the resultant 176SSAT variation can be represented by a sinusoidal function of the form, $\sin[2\pi t/170 + \phi t]$, where $\phi$ is the progressive phase shift in radians per day. As indicated in Figure 4, the phase shift over the three years (1095 days) from the beginning of 1980 to the beginning of 1983 is ~ π radians so $\phi$ ~ $\pi/1095 = 0.00287$ radians per day. This corresponds to a frequency shift from $1/170$ days$^{-1}$ to $1/170 + 0.00287/2\pi = 0.00634$ days$^{-1}$ or a shift in period from 170 days to ~ 158 days, approximately the periodicity observed by R84.

We more clearly illustrate the effect of a progressive phase shift on spectra by obtaining the frequency spectra of the observed sunspot area variations in Figures 3 and 4. Figure 5 shows the spectra of the three different components of sunspot area data. The spectra are obtained over eight years, 1977 – 1984, to provide reasonable frequency resolution by standard FFT.

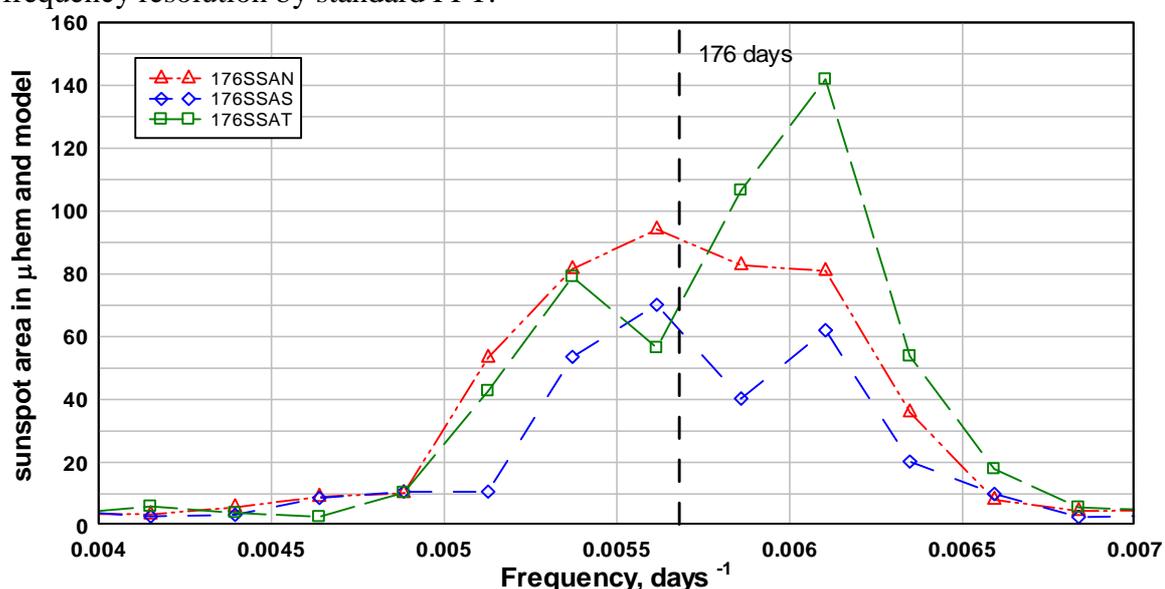

FFT 176SSAN SSAS SSAT and sum y1+y2.grf
**Figure 5.** Shows the frequency spectra of 176SSAN, 176SSAS and 176SSAT obtained over the maximum of solar cycle 21, 1977 – 1984. The shift of the spectral peak of 176SSAT to higher frequency is due, mainly, to a π phase shift in the 176SSAS variation at the beginning of 1982. This π phase shift also results in the splitting of the spectra of 176SSAS into two peaks.

The 176SSAN variation follows the secondary pulse of the model over most of the record without suffering a significant phase change and, therefore, has a broad peak centred on ~ 176 days. The 176SSAS variation follows the primary pulse of the model and the π phase change in the primary model pulse that occurs in 1982. As a result of 176SSAS following the phase change the spectrum of 176SSAS is split into two peaks centred, approximately, on 176 days. The 176 day component of total sunspot area, 176SSAT, being the sum of the 176SSAN and 176SSAS variations, suffers a progressive phase shift over the entire record, see Figure 4. This results in a shift of the major spectral peak of 176SSAT to shorter periods as evident in Figure 5. Note that the shift in spectral peak is less when the rate of phase shift is obtained over the eight year interval, 1977 to 1984, than would be the case when obtained over the four year interval, 1980 to 1984.

The above analysis explains why R84 observed a strong spectral peak at period ~ 154 days, frequency 0.0065 days$^{-1}$, 75nHz, and a weaker, longer period peak at ~ 189 days, 0.0053 days$^{-1}$, 61 nHz, in the spectra of flares. However, R84 suggested the small peak they marked A', at 61 nHz, was due to truncation of the record. Our analysis suggests the small peak at 0.0053 days$^{-1}$, 61 nHz corresponds to the peak at 0.0053 days$^{-1}$ in Figure 5 due to the residual after most of the spectral content is shifted to shorter period. The peak marked at 95 nHz marked A'' by R84 was also explained as due to truncation of the record. However, based on the more complete planetary model, Edmonds (2016b), we associate the ~ 95 nHz peak marked A'' in Figure 2, R84, with the $f_{MJ} - f_{MV} = 97$ nHz (118.5 day planetary period), and the peak, marked B at ~155 nHz in Figure 2, R84, with the $f_{MV} = 160$ nHz (72 day planetary conjunction period). This would be consistent with the observations of Hung (2007) of flare activity occurring during planet conjunctions. There are several examples of similar types of spectral splitting and shifting in sunspot area spectra reported in Edmonds (2016a).

We note that the ~154 day periodicity in flares observed by R84 is a periodicity in a derived quantity. In the present case the derived quantity is flare occurrence, a quantity resulting from the summed or averaged effects of sunspot emergence on the North and the South hemispheres of the Sun. Such derived quantities would include total sunspot number, total sunspot area, flare count, solar wind speed at Earth and geomagnetic index along with most of the other solar activity indices. These quantities are averages of effects associated with quantities, for example, SSAN and SSAS, measuring solar activity on the North or South hemispheres separately. We also note that the ~ $\pi$ radian phase changes in planetary forcing occur quite frequently, e.g., phase changes in the time variation of the primary model pulse occur at intervals of 2376 days starting at day 428 (4 March 1877) in the sunspot area record. When the phase change in the secondary pulse is included the interval between planetary phase changes is reduced to ~ 1188 days ~ 3.3 years. Therefore, we expect that a phase change in planetary forcing will often occur during the maximum part of most solar cycles. If, as we expect, sunspot emergence on one of the hemispheres is following the primary model pulse we would expect the component of sunspot emergence in this periodicity range to also follow the model phase change. As indicated above the component of sunspot emergence can respond to a model phase change with the amplitude of the component reducing to zero and re-emerging in phase with the new primary pulse. The principal effect is that components of sunspot emergence in this period range tend to occur in episodes about 3.3 years long. A further result is that, during most solar cycles, components in derived quantities such as total flare activity or total sunspot area will exhibit spectra with the major spectral peaks shifted to shorter periodicity than the fundamental planetary periodicity, as appears to be the case for solar cycle 21, discussed above. It also appears that, occasionally, North and South sunspot emergence both follow a model phase change. In such cases the frequency spectrums of both sunspot area North and sunspot area South will exhibit peaks split onto either side of the fundamental periodicity of 170 days, with one peak towards 180 days and the other peak towards 160 days. In such cases a derived quantity, e.g., total sunspot area, total sunspot number or total flare number, would also exhibit an abrupt phase change and the spectral content of the derived quantity during that solar cycle would also exhibit equal strength spectral peaks split onto either side of the fundamental planetary periodicity. For example, this type of spectral splitting is observed in the spectrum of total sunspot area for solar cycle 16 when components of both sunspot area North, sunspot area South and total sunspot area responded to a phase change in the secondary model pulse during 1926, ~ day 18422 relative to January 01, 1876. Detailed analysis of this is outside the scope of the present paper. However, we note that Gurgenashvili et al (2016) provide a useful spectral analysis, directed to the spectral range of interest here, for the derived quantity total sunspot area. Their Figure 1 demonstrates clearly that for most recent solar cycles the spectral peak in total sunspot area is shifted from the fundamental planetary periodicity, 170 days, towards shorter periodicity of ~ 160 days. It also demonstrates that in some solar cycles the fundamental periodicity at 170 days is split into two peaks, one peak towards 160 days and the other peak towards 180 days. We note that their explanation for the shifts and splits in spectra are quite different from the explanation proposed here.

The comparison between the model variation and the ~176 day component of sunspot area South was relatively straightforward as the 176SSAS component followed the primary pulse of the model with a lag of ~ 15 days and responded slowly, over about a year, to a ~ $\pi$ radian phase change in the model primary pulse by decreasing to zero amplitude and re-emerging with a near $\pi$ radian phase change. This was consistent with several examples of similar sunspot emergence observed previously, Edmonds (2016b). However, the observation that the peaks in the 176SSAN component followed the weaker secondary pulse of the model with a relatively much larger lag, ~ 40 days, was a new observation. In part, it explains why the 176SSAN component did not respond to the ~ $\pi$ radian, planetary model, phase change in 1982. The reason appears to be that, during solar cycle 21, the 176SSAN component followed the secondary pulse of the model and continued in the same phase when the secondary pulse became the primary pulse of the model. Both the 176SSAN and the 176SSAS components then became aligned with the primary pulse of the model. In part, the observation that South and North hemisphere sunspot emergence may respond consistently to different level pulses of the model, e.g. the North component responding to the secondary pulse while the South component responds to the primary pulse, explains why the periodic emergence of sunspots on the South and North hemispheres may be in quadrature, occasionally in anti-phase and occasionally in-phase and why both the Northern and Southern hemisphere components of sunspot emergence do not always follow all phase changes in the planetary model.

## 6. Conclusion

Rieger periodicity or intermediate range periodicity in solar activity has been studied extensively for over 30 years without any generally accepted explanation emerging. This is not surprising as intermediate range periodicity is exceedingly complex and intermittent. Here we present an explanation of Rieger periodicity based on a simple planetary model of sunspot area emergence that, nevertheless, generates complex time variation and frequency spectra. The analysis links the observed ~154 day periodicity in flare occurrence and in sunspot area emergence quite comprehensively to the planetary model. However, the planetary model is essentially an empirical model based on the tidal effect due to Mercury and Earth. We provide no physical mechanism whereby the resultant small periodic variations of the tidal effect on the Sun trigger sunspot emergence.

**References.**